\documentclass{aa}	
\usepackage{graphicx}
\usepackage{txfonts}
\usepackage{xspace}
\usepackage{color}

\def\mic{$\mu$m\xspace}

\def\textRef{}
\def\textRefTwee{}

\def\source{IRAS~17150$-$3224\xspace}
\begin{document}

\title{{ Micron-sized forsterite grains in the pre-planetary nebula~of~\source}}
\subtitle{ \textit{Searching for clues on the mysterious evolution of massive AGB stars }}
\titlerunning{Micron-sized forsterite grains in the outflow of IRAS~17150}


   \author{B.L. de Vries\inst{1,2}, K.M. Maaskant\inst{3}, M. Min\inst{4}, R. Lombaert\inst{5}, L.B.F.M.~Waters\inst{4,6} \and J.A.D.L. Blommaert\inst{5,7}}
   \authorrunning{B.L. de Vries et al.}

   \institute{
   		1. AlbaNova University Centre, Stockholm University, Department of Astronomy, SE-106 91 Stockholm, Sweden \\
		2. Stockholm University Astrobiology Centre, SE-106 91 Stockholm, Sweden \\
		3. Leiden Observatory, Leiden University, P.O. Box 9513, 2300 RA Leiden, The Netherlands \\
		4. Anton Pannekoek Astronomical Institute, University of Amsterdam, P.O. Box 94249, 1090 GE Amsterdam, The Netherlands \\
		5. Instituut voor Sterrenkunde, KU Leuven, Celestijnenlaan 200D, B-3001 Leuven, Belgium\\
		6. SRON Netherlands Institute for Space Research, Sorbonnelaan 2, 3584 CA Utrecht, The Netherlands \\
		7. Department of Physics and Astrophysics, Vrije Universiteit Brussel, Pleinlaan 2, 1050 Brussels, Belgium \\
              \email{bernard.devries@astro.su.se / bldevries.science@gmail.com}
             }

   \date{Received 21/08/2014 ; Accepted 17/02/2015}

 
  \abstract
   {}
   {We study the grain properties and location of the forsterite crystals in the circumstellar environment of the pre-planetary nebula (PPN) \source in order to learn more about the as yet poorly understood 
   evolutionary phase prior to the PPN.}
   {We use the best-fit model for \source of \cite{meixner02} and add forsterite to this model. We investigate different spatial distributions and grain sizes of the forsterite crystals in the circumstellar environment. We compare the spectral bands of forsterite in the mid-infrared and at 69 \mic in radiative transport models to those in ISO-SWS and Herschel/PACS observations.}
   {We can reproduce the non-detection of the mid-infrared bands and the detection of the 69 \mic feature with models where the forsterite is distributed in the whole outflow, in the superwind region, or in the AGB-wind region emitted previous to the superwind, but we cannot discriminate between these three models. To reproduce the observed spectral bands with these three models, the forsterite crystals need to be dominated by a grain size population of 2 \mic up to 6 \mic. We also tested models where the forsterite is located in a torus region or where it is concentrated in the equatorial plane, in a disk-like fashion. These models show either absorption features that are too strong or a 69 \mic band that is too weak, respectively, so we exclude these cases. We observe a blue shoulder on the 69 \mic band that cannot be explained by forsterite and we suggest a possible population of micron-sized ortho-enstatite grains. \\ We hypothesise that the large forsterite crystals were formed after the superwind phase of \source, where the star developed an as yet unknown hyperwind with an extremely high mass-loss rate ($\gtrsim$10$^{-3}$ M$_{\odot}$/yr). The high densities of such a hyperwind could be responsible for the efficient grain growth of both amorphous and crystalline dust in the outflow. Several mechanisms are discussed that might explain the lower-limit of $\sim$2 \mic found for the forsterite grains, but none are satisfactory. Among the mechanisms explored is a possible selection effect due to radiation pressure based on photon scattering on micron-sized grains.}
   {}

   \maketitle
 

%

\section{Introduction}
\source (also known as AFGL~6815 or OH~353.84$+$2.98) is a pre-planetary nebula (PPN), characterised by a stellar remnant surrounded by a previously expelled dust and gas envelope. \textRef{The infrared emission from the dust envelope of \source has been studied by \cite{guertler96}, who modelled the spectral energy distribution (SED) of that source up to wavelengths of 1.3~mm. Using a spherical geometry for the dust envelope they found that they needed either large grains or a second cold shell in order to properly fit the far-infrared emission of \source. Based on Hubble resolved images, \cite{ueta00} and \cite{meixner99} later showed that the envelope of \source is axisymmetric. \cite{meixner02} modelled both the SED and the resolved images using an axisymmetric dust distribution and \textRefTwee{amorphous grains with a minimum grain size of 0.001~\mic and a maximum size of up to sizes of several hundred micrometres. }} 

The axisymmetric envelope of \source is the material lost by \source during its asymptotic giant branch (AGB) phase. The $\text{axisymmetric}$ morphologies are also recognised in planetary nebulae themselves and these morphologies are thought to predate the PPN phase. 

\source is likely an evolutionary product of a massive ($\gtrsim$5 M$_{\odot}$) main-sequence star \citep{meixner02}. This links \source to stars that in their AGB phase stay oxygen-rich. Such massive AGB stars have mass-loss rates of the order of $\sim$10$^{-4}\,\rm{M}_{\odot}/yr$ when in their OH/IR phase \citep{vassi93}. This period of high mass-loss is often called the superwind \citep{renzini81, knapp85, bedijn87, wood92}. 

Even though the superwind of massive AGB stars was introduced so that the central star could lose enough mass to reach white dwarf masses, it is now becoming increasingly clear that the superwind is not capable of shedding enough mass. The superwind timescale turns out to be too short ($<$2000 yrs) and the number of superwind phases seems to be limited to only one. These short timescales have been shown with several methods, either based on gas and SED modelling \citep{heske90,just06,just13,decin07,just96, ches05, groenewegen12} or based on the analysis of forsterite spectral features \citep{devries14}.

These short timescales pose an evolutionary problem. The mass loss rates of massive AGB stars can go up to ${15\cdot10^{-5}\,\text{M}_{\odot}/\text{yr}}$ and even slightly higher \citep{just96, just92, schut89, groene94}. Such mass-loss rates, combined with the found timescales, gives a total mass lost in one superwind of $\sim$0.2-0.6~$\text{M}_{\odot}$ (depending on the gas over dust ratio of 100-300, \citealt{lombaert13}). Massive AGB stars need to lose several solar masses before they can leave the AGB. Therefore one would expect to see several superwind phases as extended shells around massive AGB stars and their remnants, in the same way as extended shells are seen around carbon-rich AGB stars \citep{cox12, maercker12}. It is intriguing that even though several oxygen-rich AGB stars have been observed with Herschel, none of them show any extended structure \citep{cox12}. It is likely that this is an indication that massive AGB stars have no extended shells and thus no previously ejected superwinds. It is then hard to understand how these massive AGB stars lose enough mass to evolve away from the AGB into a white dwarf.

\cite{devries14} hypothesise an intermittent phase between the PPN phase and the superwind, where the star has a mass-loss rate one or two orders higher than during the superwind. This phase of extreme mass-loss could be referred to as the \textit{hyper}wind. With hyperwind mass-loss rates of the order of 10$^{-3}$-10$^{-2}$ M$_{\odot}$/yr, several hundreds to thousands of years would be enough for the central star to evolve into a white dwarf.

The central star of a massive AGB during its superwind is already completely obscured and its SED is very red, peaking at $\sim$30 \mic. The SED of a star in its hyperwind phase would be even redder and would peak at even longer wavelengths, while its central star would still be totally obscured. So far, stars in their hyperwind phase have not been observed or identified, because the phase is so short, the superwind is still too optically thick to see the hyperwind through it, or such sources might have been wrongly classified (possibly as protostellar). However hints of the hyperwind can already be seen in the models of \source and other post-AGB sources (e.g. IRAS~16432$-$3814 in \citealt{dijkstra03} or IRAS~18276$-$1431 in \citealt{murakawa13}). \cite{meixner02} show that, at its highest, the mass-loss rate of \source must have been of the order of 8.5$\cdot$10$^{-3}$~M$_{\odot}$/yr. This makes \source a crucial object to study in the light of this evolutionary conundrum.

\begin{figure}
   \centering
   \includegraphics[width=\hsize]{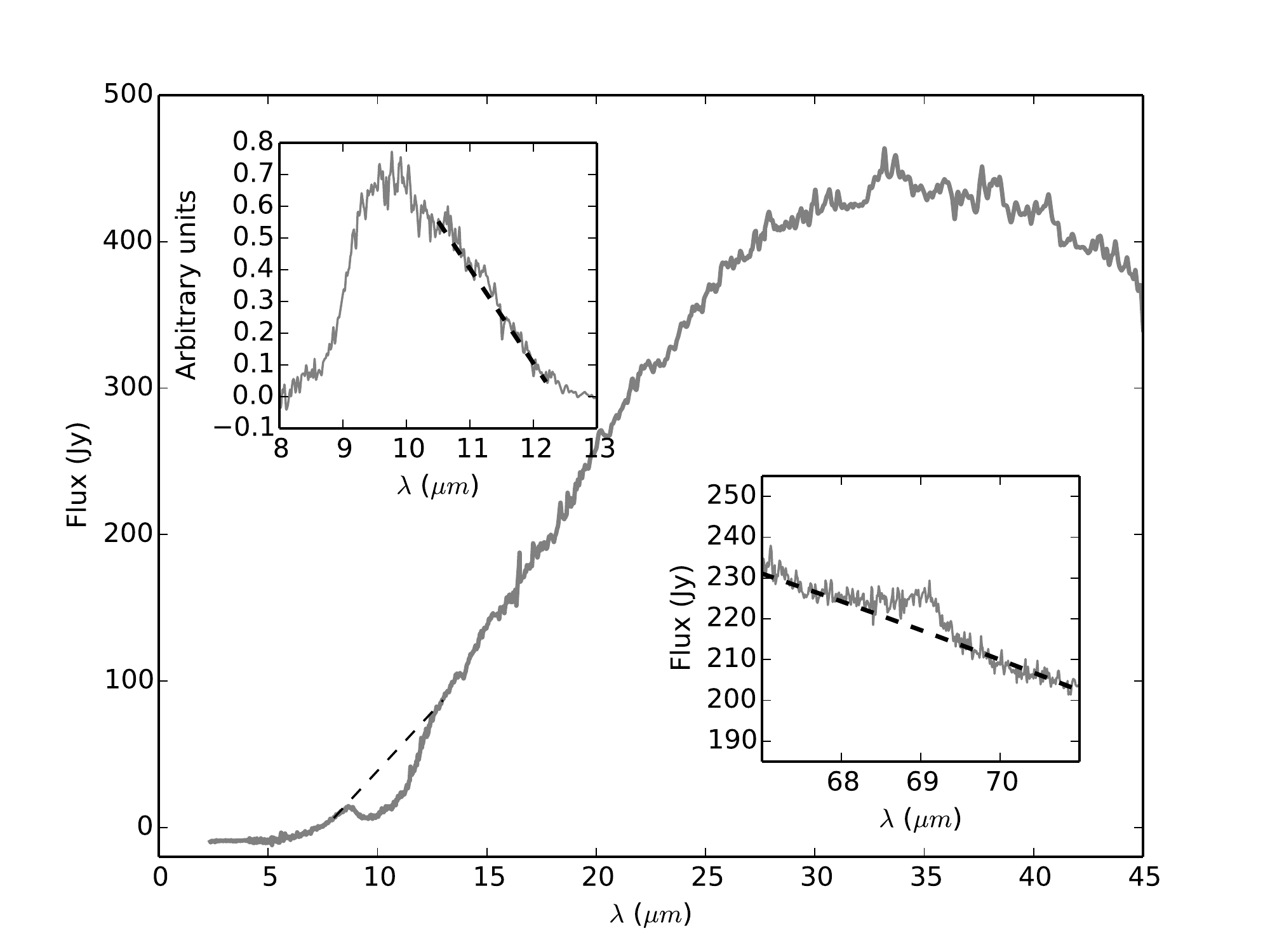}
      \caption{The main window shows the ISO-SWS spectrum of \source in grey \citep{sloan03}. The continuum constructed over the 9.7 \mic band of amorphous olivine (as explained in sect. \ref{sec:measuring}) is shown as a dashed black line. The top-left inset shows the optical depth of the 9.7 \mic absorption feature (see sect. \ref{sec:measuring}) in grey and in dashed black the continuum under the 11.3 \mic band of forsterite. In the bottom right inset we show the 69 \mic band as observed with Herschel/PACS in grey \citep{blom14} and the constructed continuum under the 69 \mic band in dashed black. }
         \label{figISO}
   \end{figure}

Another property that links \source to massive AGB stars is the presence of forsterite (Mg$_{2}$SiO$_4$) in its outflow. Massive AGB stars in their superwind phase show prominent forsterite features at among others 11.3, 33.6 and 69 \mic \citep{waters96, mol02_1, devries10, blom14, devries14}. The abundance of forsterite in these superwinds can be as high as 14\% by mass \citep{devries10, devries14}. Since the spectrum of \source does not have any convincing mid-infrared spectral features of crystalline dust species (\cite{meixner02}, see Fig. \ref{figISO}), it was thought that the outflow of \source contains no crystalline dust component, but recently \cite{blom14} reported the detection of the far-infrared spectral band at 69 \mic of forsterite. 

The study of the sharp spectral bands of forsterite is of great value since these spectral bands are dependent on many properties of the grains (see sect. \ref{sec:fo}) and thus on the outflow in general. As we will show in this work, a special combination of parameters is needed for \source in order to explain both the absence of forsterite features in the mid-infrared and a strong detection at 69 \mic. To this end we use radiative transport models.

In this paper we start by introducing the observations in sect.~\ref{sec:observations}. The behaviour of the different mid- and far-infrared bands of forsterite are described in sect. \ref{sec:fo}. Sect.~\ref{sec:method} contains the modelling strategy of this work, the radiative-transport code we use and how we study the spectral features in the model and observed spectra. We continue in sect.~\ref{sec:results} with the results and end the paper with a discussion in sect.~\ref{sec:disc} and our conclusions in sect. \ref{sec:conc}.


\section{Observations}
\label{sec:observations}
Most recently \cite{meixner02} have modelled the SED and images of \source. In their fitting they have included the B-band Hubble Space Telescope image from \cite{ueta00}, the 9.8 \mic images from \cite{meixner99} and the optical and infrared photometry of a compilation of data made by \cite{ueta00} (including data from \cite{vanderveen1989}, \cite{hu93}, \cite{kwok96}, Reddy \& Parathasarathy (1996), Meixner et al. (1999), Ueta et al. (2000), and IRAS). \cite{meixner02} were able to reproduce the SED and images of \source with an axisymmetric density distribution and by including a population of amorphous olivine grains with a size distribution \textRefTwee{from 0.001~\mic up to $\sim$200~\mic and then falling off exponentially.}

We will build on the results of \cite{meixner02} by studying the now known presence of forsterite in the outflow of \source. For this we use the ISO-SWS spectrum of \source (\citealt{sloan03}, see Fig. \ref{figISO}). The ISO-SWS spectrum shows a very smooth dust continuum and a 9.7~\mic absorption band, which are both due to amorphous silicates. As mentioned by \cite{meixner02} the ISO-SWS spectrum contains no clear indications of any crystalline silicate bands (for example at 11.3 \mic or 33.6 \mic). The presence of \textRefTwee{three weak} emission features at 33.6 \mic, $\sim$38 \mic, and $\sim$41 \mic could be argued, but this is uncertain at the end of the wavelength range of ISO-SWS. \textRef{The fact that an absorption band of amorphous silicates is seen at 9.7~\mic while no absorption band of forsterite is seen at 11.3~\mic is strikingly different from that seen in observations of massive AGB stars, the likely predecessor phase of \source.}

In contrast to the absence of mid-infrared features, the far-infrared spectrum of \source shows a clear 69 \mic band of forsterite (\citealt{blom14}, see Fig. \ref{figISO}). In the far-infrared we use the Herschel/PACS observations of the 69~\mic \citep{groenewegen11_mess, blom14} \textRefTwee{band}. The detection of the 69 \mic band in \source was reported by \cite{blom14} to be broader than could be explained from its central wavelength position. \cite{blom14} showed that a temperature gradient for the forsterite component was not enough to explain the width of the 69 \mic band of \source. In sect. \ref{sec:fo} we will explain in more detail how the band properties of the 69 \mic band depend on grain temperature and grain size.

In this work we will study the combination of the non-detection of features in the mid-infrared together with the detection of the 69~\mic band. We will also show that the broadness of the 69 \mic band as reported by \cite{blom14} is due to a shoulder on the blue side of the 69 \mic band, which cannot be due to forsterite.


\section{Forsterite features}
\label{sec:fo}
\begin{figure}
   \centering
   \includegraphics[width=\hsize]{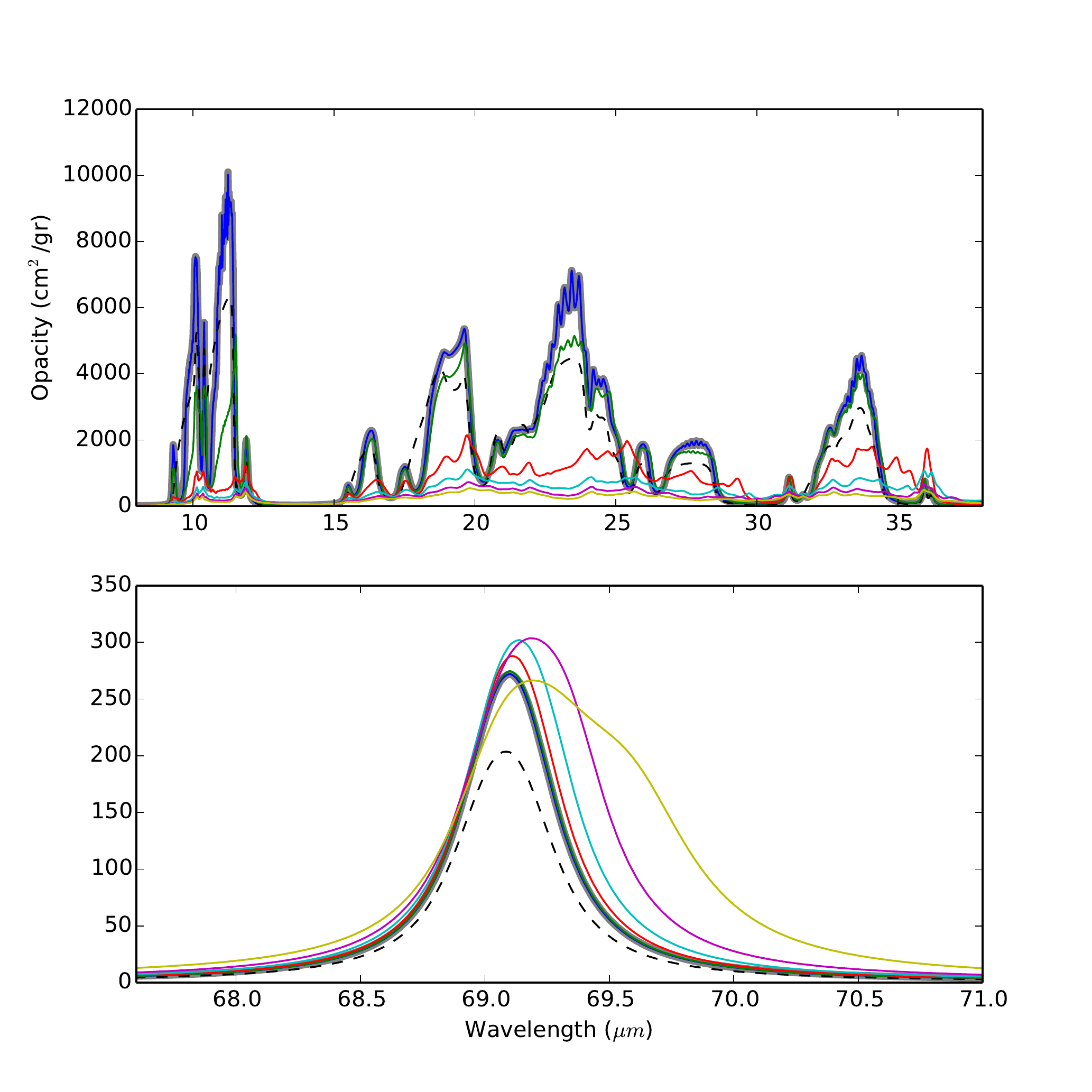}
      \caption{The effect of grain size on the mid-infrared and 69 \mic bands of 150 K forsterite grains. \textRef{The grey, blue, green, red, teal, purple and yellow curves are for forsterite with grain sizes of 0.001, 0.1, 1, 3, 5, 7 and 9 \mic, respectively. The opacities are calculated for grains with a DHS shape distribution \citep{min05}, using the optical constants of \cite{suto06}. The dashed black line are the opacities when calculated with CDE (Continuous Distribution of Ellipsoids, \citealt{BH83}) }}
         \label{figOpac}
   \end{figure}

As many different investigations have shown, the spectral features of forsterite contain a wealth of information \citep{koike03, suto06,bowey02,sturm10,mulders11,devries12,sturm13,maaskant14,blom14,devries14}. Much of this information is contained in the 69 \mic band, whose shape is strongly dependent on the grain temperature, the iron content of the crystal and\textRefTwee{,} for micron-sized grains, also on the crystal size. The band broadens and its central wavelength position shifts to the red as either the temperature increases, the iron content increases or the grain size becomes larger than $\sim$7~\mic in size \citep{koike03, suto06,sturm13, maaskant14}. Fig. \ref{figOpac} shows the behaviour of the 69 \mic band as a function of grain size. It can be seen that the effect of grain size on the 69~\mic band is small below grain sizes of 7~\mic, while above this limit the 69 \mic band broadens significantly as a function of grain size. 

The bands of forsterite in the region of the 9.7 \mic band of amorphous silicates, like the 11.3 \mic band, originate from degrees of freedom within the SiO$_4$ tetrahedral and they are therefor not sensitive to the composition of the crystalline olivine \citep{koike03}. By approximation, these bands are also not sensitive to the grain temperature \citep{zeidler12proc}. The forsterite bands in the mid-infrared (like for example the 33.6 \mic band, see Fig. \ref{figOpac}) are, compared to the 69 \mic band, mildly dependent on the grain temperature and composition \citep{koike03, zeidler12proc}. 

All spectral bands of forsterite are dependent on grain size, but at which grain size their dependence becomes significant depends on the wavelength position of the band. \textRefTwee{The grain size dependence of the bands comes from the fact that at a certain size, the interior of the grain is not sampled by the radiation field anymore}, causing the band to broaden and eventually to disappear. As is shown in Fig. \ref{figOpac}, already at grain sizes of 1~\mic the band at 11.3 \mic becomes weaker as a function of grain size. At a grain size of 1 \mic this is not yet the case, causing the band at that wavelength to still be quite insensitive to grain size. \textRefTwee{Eventually the 69~\mic band will be influenced by the grain size when the grain size becomes larger than $\sim$7~\mic.}  

\textRefTwee{Because of the grain size dependence, the peak ratios of the spectral bands are a good probe of the grain size.} This was recently shown and used by \cite{maaskant14}. They found $\sim$10 \mic forsterite crystals in the proto-planetary disk of HD~141569 by studying the relative strengths of the 69 \mic band compared to mid-infrared bands. However, before doing such a grain size analysis, a good model for the temperature and density structure of the circumstellar environment needs to be known. \textRef{For \source, a satisfactory model was derived by \cite{meixner02}, \textRefTwee{which provides us with the needed parameters to calculate} the optical depth, density and temperature of the envelope in order to study the forsterite spectral signature (see sect. \ref{sec:model}). }

\section{Method}
\label{sec:method}

\subsection{Strategy}
\label{sec:strat}
Here we discuss three ways by which the spectrum of forsterite dust would only show a 69 \mic band and no features in the mid-infrared. The first is that the material is very cold and the temperature of the material has a black-body spectrum that peaks at the far-infrared wavelengths or slightly beyond. A second option is that the medium has an optical depth that causes all the mid-infrared features to be exactly in between absorption and emission. The third option is the presence of micron-sized forsterite crystals ($\gtrsim$1 \mic ), which do not show spectral bands at short wavelengths because the grain size is significantly larger than the wavelength, but still have the 69 \mic band in the far-infrared. 

In order to find out which of the three options is the cause for the peculiar forsterite spectrum of \source, we study \textRef{forsterite features over a wide range of wavelengths, from 11.3 \mic up to the 69 \mic band.} We choose features situated at wavelengths far apart, which helps us to probe the largest possible ranges in optical depth and temperature.
The 11.3 \mic band is very sensitive to the grain size in the range between 0.1 \mic and 5.0 \mic, but not to the grain temperature and crystal composition (iron content of the crystalline olivine grains). The 69 \mic band is not sensitive to the grain size (at grain sizes $\lesssim$7 \mic), but is very sensitive to the grain temperature (for example see \citealt{suto06, koike03, sturm10, devries12}). 

The density structure of the dust around \source was determined by \cite{meixner02} (see sect.~\ref{sec:model} and Table~\ref{table:1}). We take their derived parameters for our radiative transport models and add forsterite to the model (see Table \ref{tab:abundances}). The presence of forsterite in the outflow has no effect on the density or temperature structure and the results of \cite{meixner02}, since the opacities of forsterite do not add any continuum opacity to the dust and only a small amount is needed to reproduce the 69 \mic band (see Table \ref{table:1}). 

\begin{figure*}
   \centering
   \includegraphics[width=1 \hsize]{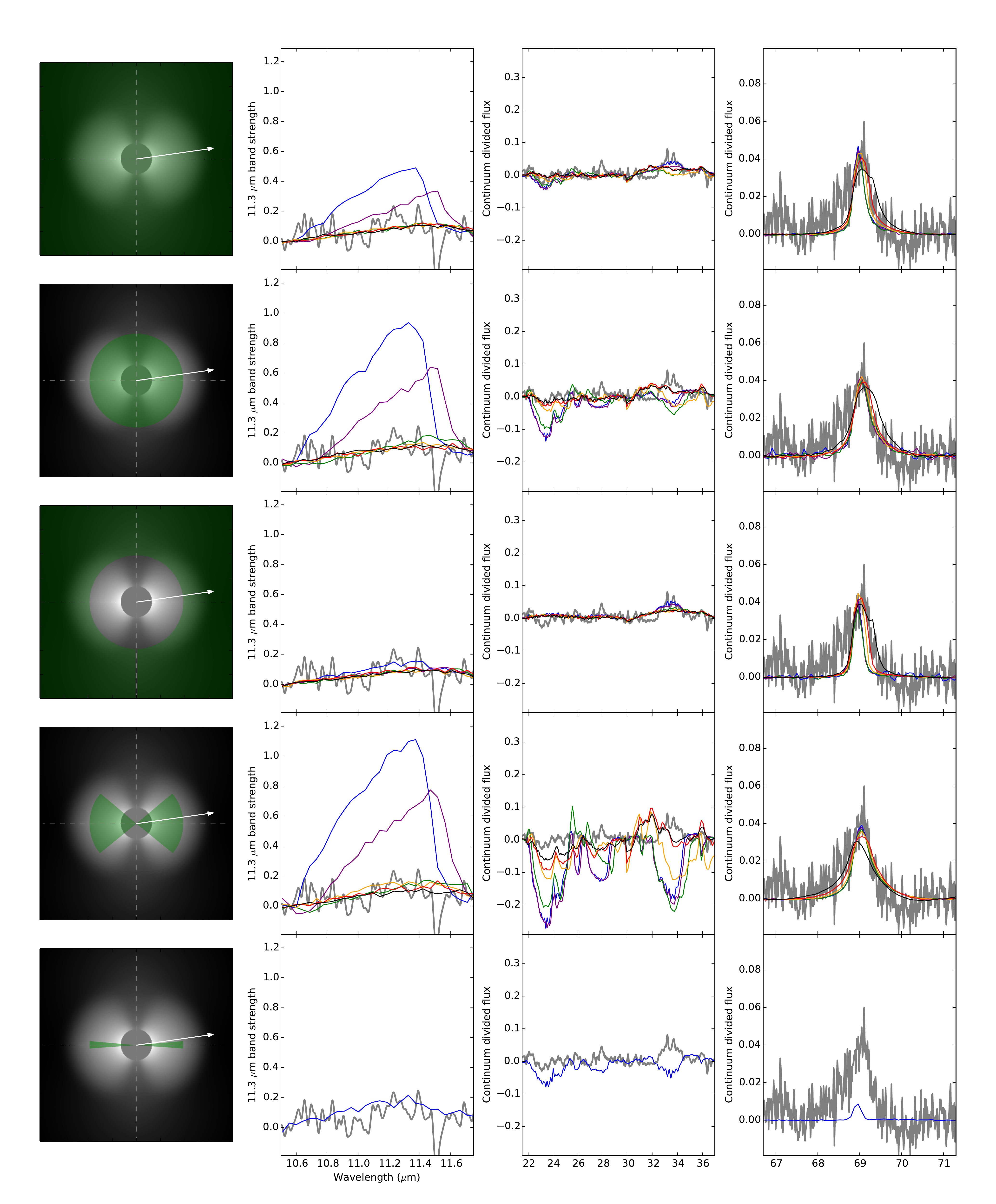}
      \caption{\textRef{The left column shows the density distributions (grey scale) and the location of forsterite (green) for the different models described in sect.~\ref{sec:strat}. The white arrow shows the inclination of the system. The second and third columns show the strength of the 11.3 \mic band and the other mid-infrared bands and the fourth column shows the flux in the 69 \mic band of forsterite, all as described in sect.~\ref{sec:measuring}. The different colours of the curves in the second, third and fourth column are for the observations (grey) and models with forsterite grains with different grain sizes: black (0.1~\mic), blue (1.0 \mic), red (2.0 \mic), orange (4.0 \mic), yellow (6.0 \mic), purple (8.0 \mic). The forsterite abundances are listed in Table \ref{tab:abundances}}.}
         \label{fig1}
   \end{figure*}

In order to extensively test all possible cases that could explain the features of \source, we explore five distributions for the forsterite, schematically depicted in \textRef{the left column} of Fig. \ref{fig1}:
\begin{itemize}
\item Model \textit{Whole}: forsterite follows the same density distribution as the amorphous dust and is present in both the superwind and the low mass-loss rate AGB-wind previous to the superwind.
\item Model \textit{Superwind}: forsterite is only present in the superwind (R $ < $ R$_{\rm{SW}}$).
\item \textRef{Model \textit{AGB-wind}: forsterite is only present in the outflow that happened previous to the superwind (R $ > $ R$_{\rm{SW}}$).}
\item Model \textit{Torus}: forsterite is only present in the densest part of the superwind (R $ < $ R$_{\rm{SW}}$, $\theta$ $>$ 50$^{\rm{o}}$)
\item Model \textit{Disk}: forsterite is only present in a disk-like region in the superwind (R $ < $ R$_{\rm{SW}}$, $\theta$ $>$ 85$^{\rm{o}}$)
\end{itemize}
Here $\theta$ is the angle from the pole and R is the radial distance from the centre of the star. For all models, in the region where forsterite is present in the outflow, it follows the density distribution of the amorphous silicates (see Eq. \ref{eq: dens}).

For these \textRef{five} models we test a grid of six grain sizes for forsterite: $\le$0.1, 1.0, 2.0, 4.0, 6.0 and 8.0 \mic. \textRef{We do not consider smaller grain sizes for forsterite since the absorption opacities do not change when the size is decreased from 0.1 \mic (see Fig. \ref{figOpac})}. For these different grain sizes we run a grid of models with different forsterite abundances from 0.5\% up to 25\% \textRef{(for the \textit{Disk} model only did we increase the abundance to 90\% in order to get a signal at 69 \mic)}. In these grids we search for the models with abundances that correctly predict the strength of the observed 69 \mic band. \textRefTwee{The abundances found} for the different models are listed in Table \ref{tab:abundances}. Among the models that reproduce the 69~\mic band we search for \textRefTwee{one} that also explains the absence of the \textRef{mid-infrared bands} in the spectrum of \source.

\begin{table}
\caption{Best-fit model parameters of \cite{meixner02}}             
\label{table:1}      
\centering                          
\begin{tabular}{l r r r r r r}        

\multicolumn{7}{c}{\cite{meixner02} fit parameters} \\
\hline                        
\hline                                   
L$_{\rm{star}}$ ( L$_{\odot}$)			& \multicolumn{6}{c}{27,200} \\
T$_{\rm{star}}$ (K) 					& \multicolumn{6}{c}{5200} \\
R$_{\rm{star}}$ (R$_{\odot}$)			& \multicolumn{6}{c}{201}\\
Distance (kpc)						& \multicolumn{6}{c}{3600}\\
R$_{\rm{in}}$ (AU)					& \multicolumn{6}{c}{648.4}\\
R$_{\rm{SW}}$ (AU)					& \multicolumn{6}{c}{1944.4}\\
R$_{\rm{out}}$ (AU)					& \multicolumn{6}{c}{$64\cdot10^3$}\\
v$_{\rm{exp}}$ (km/sec)				&\multicolumn{6}{c}{15} \\
Inclination							& \multicolumn{6}{c}{82.0$^{\circ}$} \\
Total dust mass (M$_{\odot}$)			& \multicolumn{6}{c}{$4.6\cdot10^{-2}$}  \\
A$_{\rm{meix}}$ 					& \multicolumn{6}{c}{159} \\
B$_{\rm{meix}}$ 					& \multicolumn{6}{c}{2.0} \\
C$_{\rm{meix}}$ 					& \multicolumn{6}{c}{1.5} \\
D$_{\rm{meix}}$ 					& \multicolumn{6}{c}{1.0} \\
E$_{\rm{meix}}$ 					& \multicolumn{6}{c}{4.0} \\
F$_{\rm{meix}}$ 					& \multicolumn{6}{c}{1.5} \\
Am. grain size 				& \multicolumn{6}{c}{0.001 - 200 \mic} \\
Am. grain size slope			& \multicolumn{6}{c}{\textRefTwee{-3.5}}\\
\\
\label{tab:modelparam}
\end{tabular}
\end{table}

\begin{table}
\caption{Forsterite abundances used for the models in Fig. \ref{fig1}}             
\label{table:1}      
\centering                          
\begin{tabular}{l r r r r r r}        

\multicolumn{7}{c}{Forsterite abundance (\%, see sect. \ref{sec:strat})} \\
\hline                        
\hline                                   
Grain size	(\mic):			& $<$0.1 	& 1 		& 2 		& 4 		& 6 		& 8 \\
\hline
\textit{Whole} 			& 1.3 	& 1.3		& 1.3 	& 1.6 	& 1.8 	& 2.1 \\
\textit{Superwind}		&  3.8	& 3.8 	& 3.8 	& 3.8	 	& 3.8 	& 3.8 \\
\textit{AGB-wind}		& 1.6 	& 1.6 	& 1.6 	& 2.6 	& 3.4 	& 4.8 \\
\textit{Torus} 			& 10.6 	& 10.6 	& 10.6 	& 11.5 	& 12.3 	& 11.9 \\
\textit{Disk} 			& 90.0 	& - 	& - 	& - & - & - \\

\label{tab:abundances}
\end{tabular}
\end{table}

\subsection{Models}
\label{sec:model}

\textRef{\cite{meixner02} fitted the SED up to 1.1 mm as well as the optical and near-IR resolved images of \source. It is beyond the scope of this work to improve on this already satisfactory model, and we adopt its parameters for the central star and the circumstellar environment (see Table \ref{tab:modelparam}). In Sect. \ref{sec:base}, we do test the robustness of our results when uncertainties are introduced in the total dust mass and the stellar effective temperature determined by \cite{meixner02}. \cite{meixner02} concluded that the best fit was obtained with an equatorially }enhanced density distribution,
\begin{eqnarray}
\label{eq: dens}
\rho(R, \theta) = \rho_{\rm{in}} \left( \frac{R}{R_{\rm{in}}} \right) ^{-B (1+C \rm{sin}^{F} (\theta) (\rm{e}^{-(R/R_{\rm{SW}})^D} / \rm{e}^{-(R_{\rm{in}}/R_{\rm{SW}})^D})) } \\
\times [1+A(1-\rm{cos}(\theta))^F (\rm{e}^{-(R/R_{\rm{SW}})^E} / \rm{e}^{-(R_{\rm{in}}/R_{\rm{SW}})^D})] \nonumber,
\end{eqnarray}
with $R_{\text{in}} < R < R_{\rm{SW}}$; $R$ is the distance from the centre of the central star and $\theta$ is the angle from the polar direction ($\theta=0$); $R_{\rm{in}}$ and $\rho_{\rm{in}}$ are the inner radius and the density at the inner radius of the dust shell; and $R_{\rm{SW}}$ is the superwind radius outside of which the outflow is just spherically symmetric. \cite{meixner02} create a time-independent outflow (in all directions) by using an $r^{-2}$ density distribution ($B=2.0$).

The parameters of the best-fit of \cite{meixner02} are shown in Table \ref{tab:modelparam}. \textRef{The total dust mass determined by \cite{meixner02} is 4.6 $\cdot$ 10$^{-2}$ M$_{\odot}$. This mass is constrained by the shape and peak of the SED and the optical depth in the 9.7~\mic absorption band of amorphous silicates, in addition to reproducing the resolved images. The best fit of \cite{meixner02} required amorphous grains with \textRefTwee{an exponential cut-off at} 200 \mic in size in order to reproduce the mm fluxes, as well as small amorphous grains (\textRefTwee{down to 0.001} \mic) to fit the near and mid-infrared part of the SED properly. For the central star we use a temperature of 5200 K. As shown by \cite{sanchez08} this temperature might be too low. Changes in the temperature of the central object will change the dust temperature in the inner region of the circumstellar environment. However, as shown in sect. \ref{sec:base} and by \cite{devries12}, this will have no significant effects on the modelling of the dust in the case of very high optical-depth environments, because then the inner parts of the outflow are not directly observed}

We have reproduced the best-fit model of \cite{meixner02} using the radiative transport code MCMax \citep{min09}. \textRef{The MCMax code has been widely and successfully applied to model observables in a variety of environments \citep{mulders11, min13, lombaert12, lombaert13}. We refer to these papers for a full description of the features of MCMax. MCMax first computes the radiative equilibrium temperature stratification throughout the circumstellar environment. The temperature structure is calculated for the different grains present in the circumstellar environment separately, so no thermal coupling between different types of dust grains is assumed. \textRefTwee{For the size distribution of the amorphous grains, the opacity of grains with different sizes are weighted and combined}. Using the resulting temperature structure, spectra can be obtained by ray-tracing. Dust scattering can be treated in an angle-dependent way, or, to speed up the computations \citep{min09} by using isotropic scattering. In this work we use isotropic scattering.}

We use the same size distribution and optical constants as \cite{meixner02} for the amorphous olivine (MgFeSiO$_4$), namely those of \cite{dorschner95}, and we extend the model of \cite{meixner02} by including forsterite grains. MCMax uses temperature-dependent opacities, meaning the model correctly predicts the 69 \mic band shape depending on the temperature of the forsterite grains \citep{mulders11}. For crystalline olivine we use the optical constants of forsterite from \cite{suto06}, which are available for grain temperatures of 50, 100, 150, 200 and 295 K. 

The opacities of the dust grains are calculated using the Distribution of Hollow Spheres (\cite{min03}, f$_{\text{max}}$=1.0). \textRef{Other particle shape models used in the literature are spherical grains and the Continuous Distribution of Ellipsoids (CDE, \citealt{BH83}). We do not consider spherical grains because it has been shown that for crystalline dust species this model predicts spectral features with shapes and wavelength positions that do not compare with spectra of any astronomical object \citep{min05}. \cite{min05} also show that the DHS model reproduces the properties of forsterite spectral bands in spectra of astronomical objects very well. We also do not consider CDE grains for two reasons. The first is that it predicts very similar spectral features in shape and position as DHS (see Fig. \ref{figOpac}). A difference between spectral features of CDE compared to 0.1~\mic DHS grains is that they are slightly weaker. Second, CDE is only valid for particles small compared to the wavelength of the radiation considered and does not allow us to model grains of different sizes, which is essential for this work.}



   \begin{figure}
   \centering
   \includegraphics[width=\hsize]{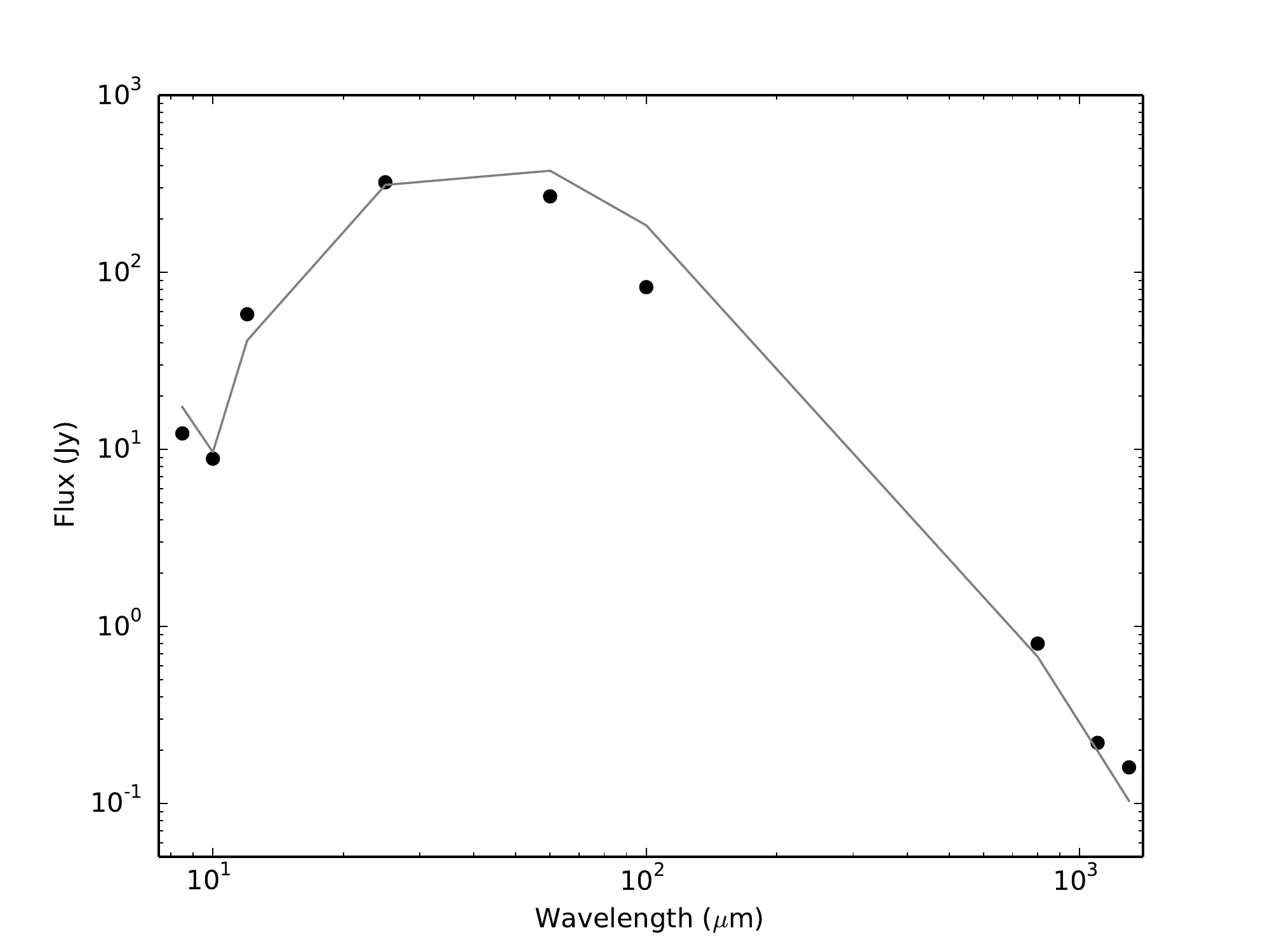}
      \caption{\textRef{The spectral energy distribution of \source in black dots. The 8.5 and 10.0 \mic points are from \cite{ueta00}, the 12, 25, 60, and 100~\mic points from IRAS, the 800.0 and 1100.0 \mic points from \cite{hu93} and the 1.3 mm from \cite{guertler96}. In grey is our model based on the parameters of \cite{meixner02}. } }
         \label{figSED}
   \end{figure}

\subsection{Extracting the forsterite spectral features}
\label{sec:measuring}
A possible 11.3 \mic band would be situated on top of the red shoulder of the 9.7~\mic band of amorphous silicate, which is in absorption (see the inset of Fig. \ref{figISO}). The 11.3 \mic band has been extensively studied in massive AGB stars by \cite{devries10} and \cite{devries14}. They study the ratio of the optical depth of forsterite relative to that of the amorphous continuum dust in the 9.7 \mic band. This ratio (here called the strength \textit{S} of the 11.3~\mic band) is interesting because in the optically thick limit it is proportional to the opacities (and abundance) of forsterite,
\begin{equation}
\text{S} = \frac{ \tau_{\text{fo}} }{ \tau_{\text{all}} } \propto \frac{ A_{\text{fo}} \cdot \kappa_{\text{fo}} }{ \kappa_{\text{all}} },
\label{eq: S113}
\end{equation}
where $\tau_{\text{of}}$ and $\tau_{\text{all}}$ are the optical depth of forsterite only and the optical depth of all dust species together, respectively, and $A_{\text{fo}}$ is the abundance of forsterite.
Since the circumsteller environment of \source is very dense and the optical depth is much larger than one, equation \ref{eq: S113} holds. This tells us that if there is forsterite in the line of sight to the star and the crystals do not approach a grain size of one micron, an absorption band of forsterite must be seen on the red shoulder of the 9.7 \mic band of amorphous olivine. If the forsterite grains are a micron or larger in size the strength $S$ will quickly go to zero.

We briefly describe how we measure the strength of the 11.3~\mic band (eq. \ref{eq: S113}) and refer to \cite{devries10} for further details. In order to calculate the $\tau_{\text{all}}$ and $\tau_{\text{fo}}$ we first construct a continuum over the 9.7~$\mu$m band of amorphous olivine (as shown in Fig. \ref{figISO}). The optical depth in the 9.7~$\mu$m band is calculated using $\tau = -\text{ln}(F/F_{\text{cont}})$, where $F_{\text{cont}}$ is the flux level of the continuum constructed over the 9.7~$\mu$m band of amorphous silicate. The optical depth in the 9.7~$\mu$m band due to forsterite ($\tau_{\text{fo}}$) and due to the other dust species ($\tau_{\text{all}}$) is now determined by constructing a continuum in $\tau$-space, under the 11.3~$\mu$m band (see again Fig. \ref{figISO}). For the 69 \mic band, we follow the approach of \cite{blom14} by fitting the continuum with a linear function (see Fig. \ref{figISO}). \textRef{The mid-infrared bands are extracted in the region of 21 to 37 \mic by fitting a spline to the continuum points. The continuum points are taken at (all in \mic) 21.7-21.9; 26.5-26.7; 29.4-31.8; 36.7-37.8.}

   \begin{figure}
   \centering
   \includegraphics[width=\hsize]{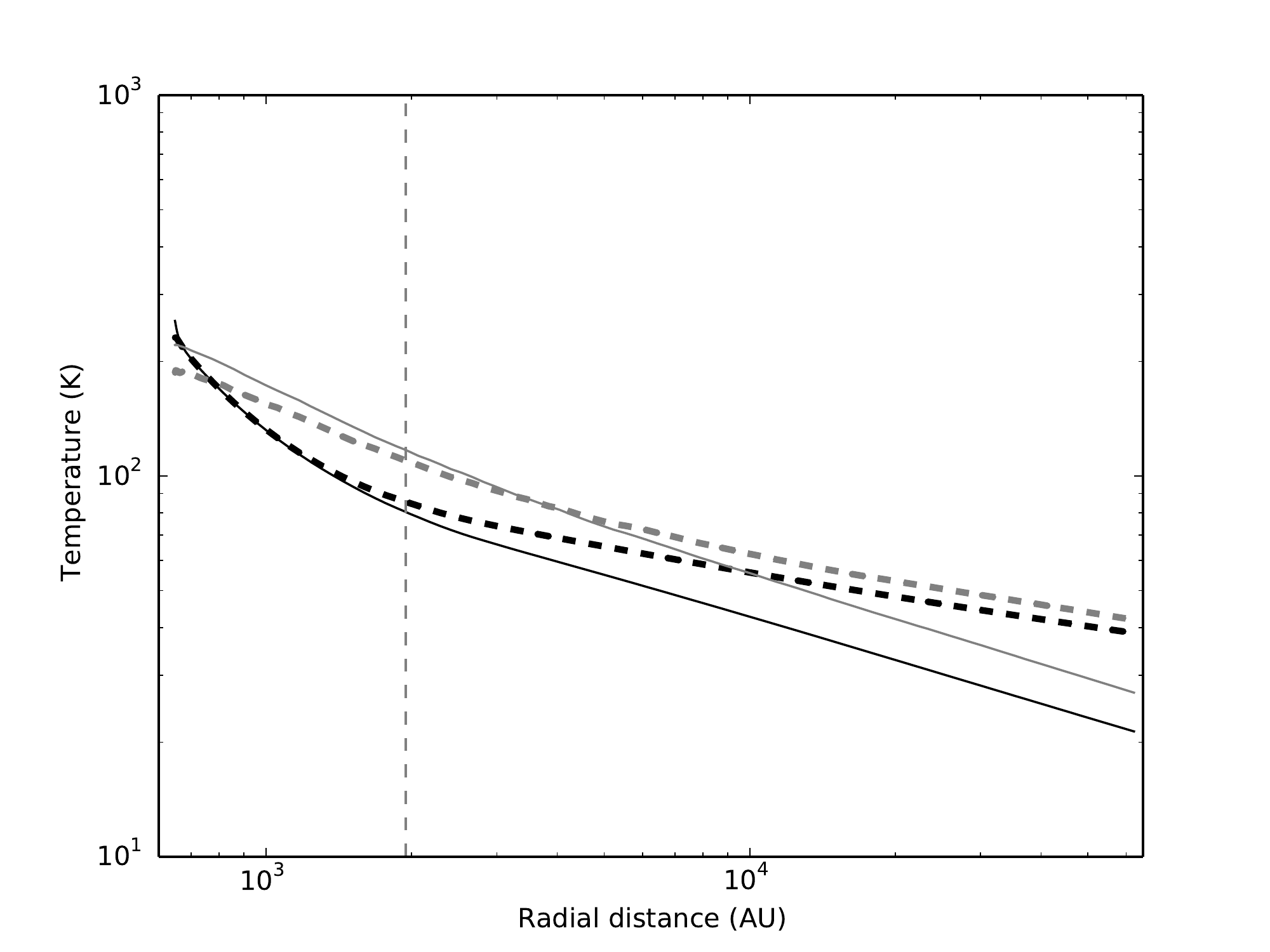}
      \caption{ \textRef{Temperature gradient of the amorphous silicate (solid line) and the forsterite (dashed line) grains for the \textit{Whole} model (see sect. \ref{sec:strat}). \textRefTwee{Black lines give the temperature structure over the equator, while the grey gives it at the pole.} The grey vertical dashed line indicates the transition between the low-density AGB wind and the superwind at the radius of R$_{\rm{SW}}=$ 1944 AU}.}
         \label{figT}
   \end{figure}

\subsection{Base model and testing the results of \cite{meixner02}}
\label{sec:base}

We find that our model SED, like the model SED of \cite{meixner02}, \textRefTwee{matches the observed photometry well enough to study the spectral bands of forsterite} \textRef{(see Fig.~\ref{figSED}). Similar to the model of \cite{meixner02}, we slightly over-predict the flux at 100 \mic, but due to the large amorphous grains added by \cite{meixner02} the fluxes around 1.0 mm are well reproduced. It is outside the scope of this work to try and improve the fit to the SED. }

\textRef{The spectral features we obtain by adding forsterite to the model with the best-fit parameters of \cite{meixner02} are shown in Fig. \ref{fig1}. We see that the model reproduces the observed 69 \mic band of \source well in strength, but its wavelength position is too blue and too narrow. This means that for this model the forsterite grains are too cold and/or too small. The mid-infrared bands of forsterite in the model spectrum are almost completely absent, similar to what is observed. The strength of the 11.3~\mic band on the other hand is too high in our model compared to the absence of any 11.3~\mic band in the observed spectrum. In section \ref{sec:results} we will show what happens to the spectral features of forsterite if the forsterite grain size and geometrical distribution are varied.}

\textRef{Since different values for the total dust mass in the outflow and the central temperature of the star have been reported \citep{sanchez08, guertler96} we want to test if a difference in these parameters has an effect on the spectral features of forsterite in the spectrum of \source. We investigate this by varying the central temperature and dust mass by $\pm$10\%. We found this has no significant effect on the spectral features. We therefor conclude that the fit of \cite{meixner02} and the parameters they obtained are of sufficient quality to reproduce the optical depth and temperature structure of the circumstellar environment in order for us to study the crystalline dust component in the outflow.}


\textRef{In Fig. \ref{figT} we show the equatorial and polar temperature gradient for both the amorphous dust and forsterite that we obtain with our model. We find that the temperature of the \textRefTwee{amorphous dust} component, at the inner radius of the dust shell, reaches $\sim$220~K, which is the same value as found by \cite{meixner02}. The temperature gradients also show that \textRefTwee{at the equator and }within the superwind (R$_{\rm{SW}}<$ 1944 AU) the optical depth is so high that the temperature of the dust is determined by the local radiation field of the dust. This means that both the amorphous dust and the forsterite have the same temperature. Outside the superwind the optical depth drops and the temperature structure of the amorphous dust and the forsterite starts to differ and the forsterite becomes slightly warmer than the amorphous dust. The forsterite is warmer than the amorphous dust because 1) the amorphous dust can cool more efficiently than forsterite since it has higher far-infrared opacities and 2) since both dust species are now heated by the radiation field of the inner dust shell. This shell radiates most of its energy in the mid-infrared, where forsterite has a higher overall absorption opacity. \textRefTwee{The optical depth at the pole is lower than at the equator and the forsterite is slightly cooler than the amorphous dust.}}


\section{Results}
\label{sec:results}


\textRef{The mid-infrared and 69 \mic bands of forsterite we find for the observations and models are shown in Fig. \ref{fig1}. In this section we will discuss the forsterite features of the five different models introduced in section \ref{sec:strat}.}

\subsubsection*{Whole model}
\textRef{In Fig. \ref{fig1} it can be seen that the 11.3 \mic band in the \textit{Whole} models for forsterite grain sizes of $\le$0.1 and 1.0 \mic are too strong. Only when the grain size is increased to $\ge$2~\mic does the 11.3 \mic band become weak enough and compare well with the observed absence of the 11.3 \mic band. The mid-infrared spectral bands do not strongly show in the models for any forsterite grain size. The 69 \mic band for models with forsterite grain sizes below 6 \mic have a wavelength position that is bluer and a shape that is more narrow than the observed band. The model with a forsterite grain size of 6.0 \mic reproduces the observed 69 \mic band well, while the 8 \mic forsterite grain size model has a too broad 69 \mic band. }

\subsubsection*{Superwind model}
\textRef{Similar to the \textit{Whole} model, the 11.3 \mic bands of the \textit{Superwind} models are only weak enough for the models with forsterite grain sizes of $\ge$2 \mic. The \textit{Superwind} models with grain sizes of $\le$2 \mic show absorption bands in the mid-infrared, especially at $\sim$ 23.5~ \mic. Only models with forsterite grain sizes of $\ge$4 \mic have weak enough mid-infrared bands to reproduce the observed spectrum. The 69 \mic bands of the \textit{Superwind} models reproduce the observed band very well, except for the 69~\mic band in the model with 8 \mic sized forsterite grains. This model has a 69~\mic band that is too broad.}

\subsubsection*{\textRefTwee{AGB-wind model}}
\textRef{For this model the forsterite is so cold that no spectral features are seen at 11.3~\mic and in the mid-infrared for any forsterite grain size. Because the forsterite is so cold, the 69~\mic band is also very narrow and blue shifted and the models with forsterite grain sizes of $\le$4 \mic do not reproduce the 69~\mic band well. At 8~\mic forsterite grain sizes the 69~\mic band is too broad, so only the 6~\mic forsterite grain size model reproduces the 69~\mic band well. The jagged feature on the red shoulder of the 69~\mic band for the 8~\mic forsterite grain size model occurs because the 69~\mic band develops a mild double-peaked shape at grain sizes of $\ge$8~\mic and low temperatures (Fig. \ref{figOpac}).}

\subsubsection*{Torus model}
\textRef{Similar to the \textit{Whole} and \textit{Superwind} model the \textit{Torus} model with forsterite grain sizes of $\le$2 \mic have a too strong 11.3~\mic band. In the mid-infrared the \textit{Torus} models show absorption bands (especially at 23.5 \mic) for all the forsterite grain sizes, except possibly for the 8 \mic model where the absorption features are weak enough to reproduce the observed absence of mid-infrared bands. All but the 6 and 8 \mic forsterite grain size models reproduce the 69 \mic band well for the \textit{Torus} model.}

\subsubsection*{Disk model}
\textRef{We had to increase the forsterite abundance for this model to 90\% in order to see a signal at 69 \mic. It can be seen that even at this abundance the strength of the 69 \mic band \textRefTwee{cannot} be reproduced. }

\subsection*{Blue shoulder on the 69 \mic band}
\begin{figure}
   \centering
   \includegraphics[width=\hsize]{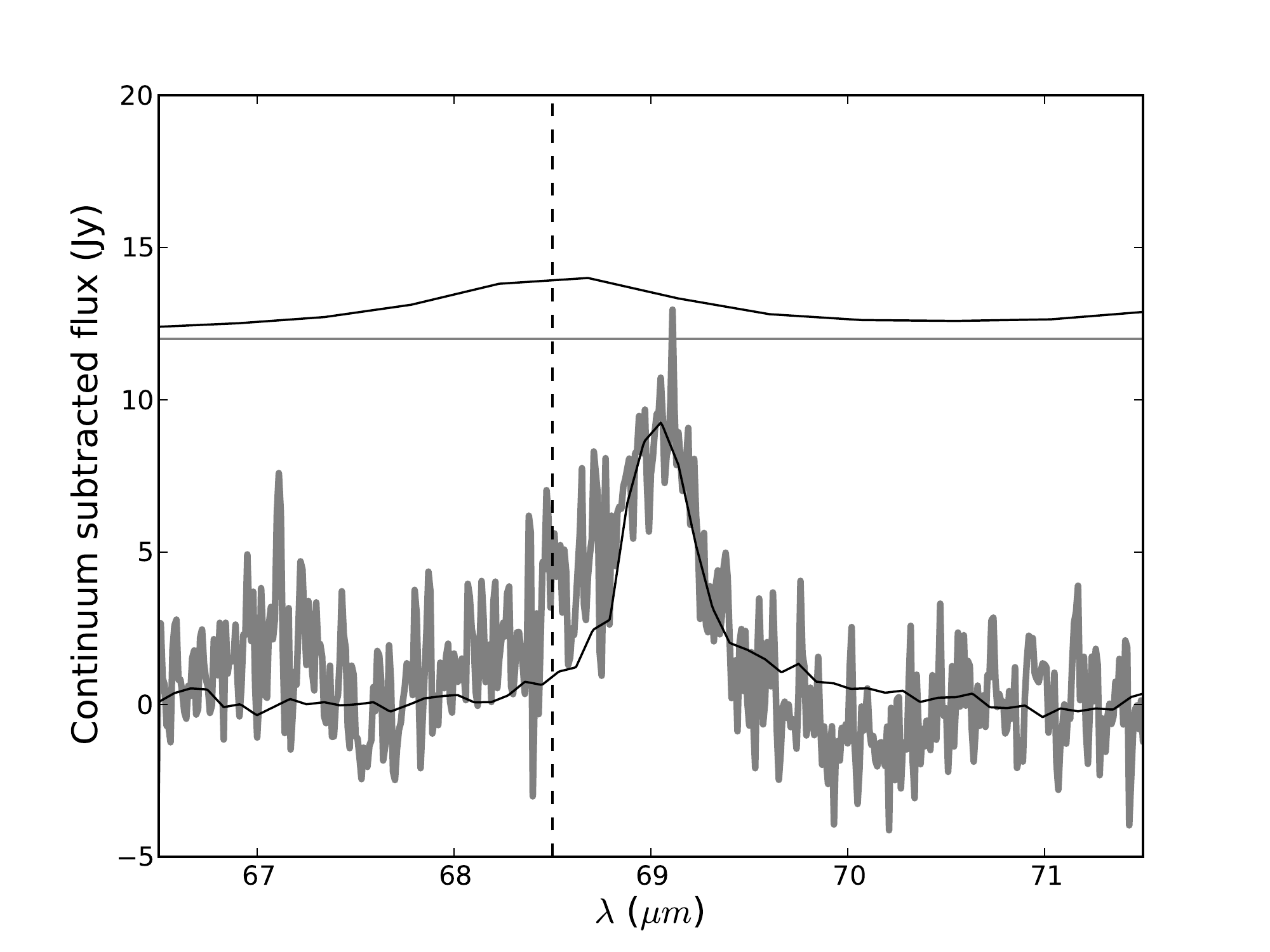}
      \caption{Shown in grey is the observed, continuum subtracted 69 \mic band and over plotted in black a radiative transport model (\textit{Whole} model, forsterite grain size of 4 \mic). Shifted to the horizontal grey line are shown the scaled absorption opacities of ortho-enstatite \citep{chihara02} in black. Ortho-enstatite has a resonance around 68.5 \mic, indicated with a dashed black vertical line.}
         \label{figEn}
   \end{figure}

The centre and red shoulder of the 69 \mic band of \source can be well reproduced by some of our models, while the small shoulder on the blue side is not (see also Fig.~\ref{figEn}). The blue shoulder being located at roughly 68.5 \mic means it is too far to the blue side for it to be caused by a cold forsterite component \citep{suto06, koike03}. We suspect that this blue shoulder might be responsible for what caused \cite{blom14} to note that the 69 \mic band is too broad for its central wavelength position. An explanation can be the presence of ortho-enstatite (MgSiO$_3$, \citealt{chihara02}), which has a resonance at 68.5 \mic. See Fig. \ref{figEn} for a comparison of the blue shoulder and the opacities of ortho-enstatite. Ortho-enstatite has another resonance at 72.5 \mic, but this is in the middle of the two Herschel/PACS bands, making it difficult to study the presence of this spectral feature. Several resonances also occur in the mid-infrared, but they are not detected in the ISO-SWS spectrum. However, if the ortho-enstatite grains are of the same size as the forsterite grains, the mid-infrared bands would be suppressed as well. Since no other resonances are known to us around 68.5 \mic, we find it likely that ortho-enstatite is present in the outflow of \source with a grain size of a few micrometres.

\subsection*{Summary}
\textRef{We can exclude the \textit{Disk} model since we cannot reproduce the 69 \mic band with an abundance as high as 90\%. We can also exclude forsterite grains of size $\le$2 \mic for all models. For the \textit{Whole}, \textit{Superwind} and the \textit{Torus} model this is based on the strength of the 11.3 \mic band. For the \textit{AGB-wind} model this is based on the wavelength position of the 69 \mic band. }

\textRef{We can exclude forsterite grain sizes of $\ge$8 \mic because for all models the 69 \mic band becomes too broad. It is difficult to find a \textit{Torus} model that does not have strong mid-infrared bands and we therefore find the \textit{Torus} model unlikely. For the \textit{Whole}, \textit{Superwind} and \textit{AGB-wind} case, satisfactory models can be found for forsterite grain sizes between 2~\mic and 6~\mic.}

   \begin{figure}
   \centering
   \includegraphics[width=\hsize]{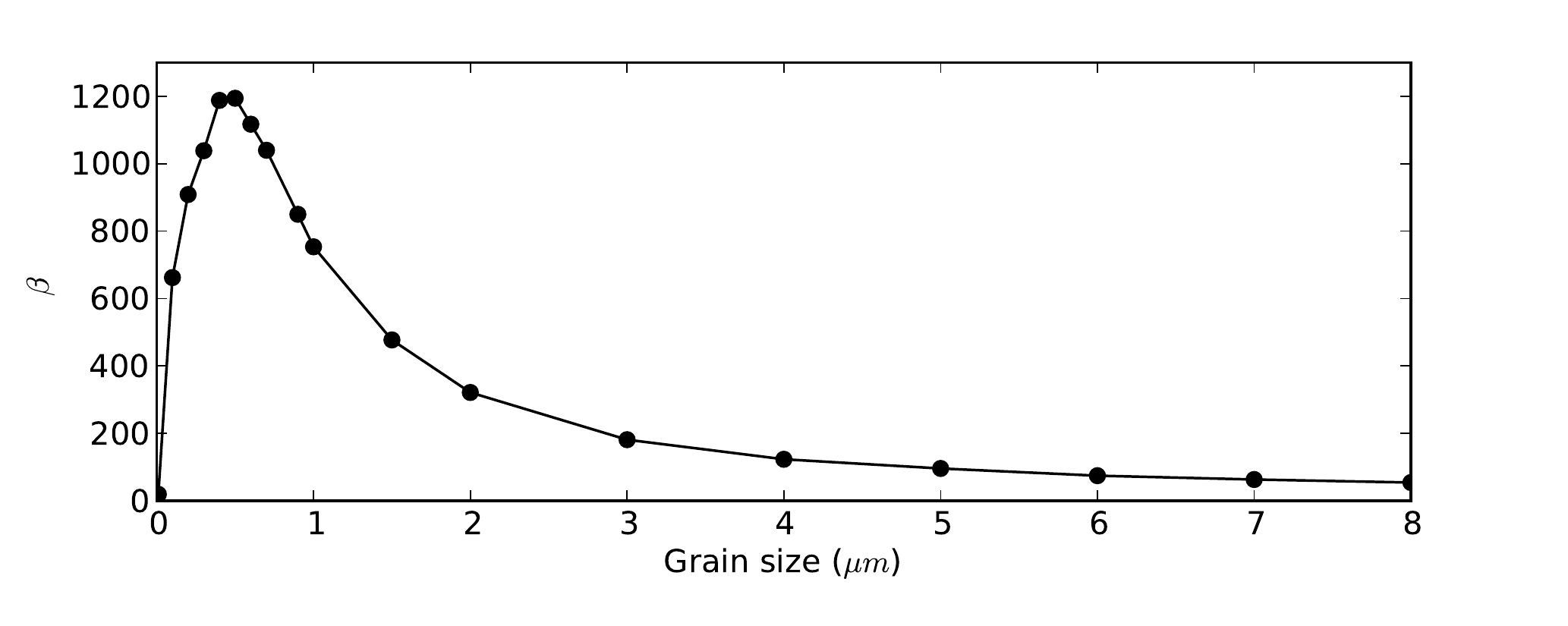}
      \caption{Ratio of the radiation pressure over the gravitational force ($\beta$) for forsterite grains of different grain sizes. }
         \label{figBetas}
   \end{figure}

\section{Discussion}
\label{sec:disc}
\textRef{All of our models indicate micron-sized forsterite grains not smaller than 2 \mic and not larger than 6 \mic. This shows that the forsterite is still significantly smaller than the largest amorphous grains ($\sim$200~\mic). From dust condensation theory this difference can be understood, since the formation and growth of crystalline solids requires more specific conditions \citep{tielens98, gailsedl99, sogawa99}. A high temperature (above $\sim$ 1000 K) and pressure together with sufficient cooling times are needed to form a crystalline solid. Otherwise an amorphous solid is formed instead. In the outflows of evolved stars conditions for crystalline dust formation are only met in some cases, close to the central star in the inner part of the outflow. Because of this it can be understood that only small amounts of small crystals can be formed. This in contrast to amorphous solids, which can be formed and grown in larger regions of the outflow.}

The fact that \cite{devries14} can model consistently the 11.3 \mic and 69 \mic bands in massive AGB stars, means the forsterite crystals in the superwinds of these objects are sub-micron in size. We then find a large difference in forsterite grain size in superwinds of AGB stars and in the outflow of the PPN \source. At this moment there are no reports of large amorphous silicate grains in superwinds and the multitude \textRefTwee{of} SED fits to superwind spectra in the far-infrared should have found large grains if they were present \citep{just06,just13}. For the massive AGB star OH~26.5+0.6, \cite{groenewegen12} finds an amorphous grain size between 0.15 \mic and 0.25 \mic. 

The high mm-fluxes of \source together with our analysis of the forsterite spectral features show that there is an increase in grain size for both the amorphous and crystalline component in-between the superwind and the PPN phase. If there indeed exists a hyperwind, the conditions during it must lead to very efficient grain growth and crystallisation. We associate a very high density with the hyperwind and indeed dust condensation theory suggests that the density is critical in grain growth and the formation of crystalline silicates \citep{tielens98, gailsedl99, sogawa99}. Besides the density, the temperature is also critical for the formation of crystalline solids. For olivine the temperature needs to be higher than its glass-temperature of $\sim$1000 K, for it to condense into crystalline forsterite. Since we see no indication of a disk-like structure, the formation of the large forsterite crystals we see around \source likely happened in the hot dust condensation region close to the star during the hyperwind, \textRefTwee{although we cannot exclude the \textit{AGB-wind} model}.

The firm lower-limit we find for the forsterite grain size is a challenge to explain. If the formation of crystalline material is efficient during the hyperwind, why is the forsterite grain size dominated by micron-sized crystals and why is there no population of sub-micron crystals? One possible mechanism we explored is based on the wind driving mechanism in oxygen-rich AGB outflows. The outflow of AGB winds is thought to be initiated by pulsations, which  pushes gas to distances from the star where it can condense into dust grains \citep{Wickramasinghe66, Gehrz71, Sedlmayr94, HO03}. Then radiation pressure on the most refractory dust grains drives the outflow to escape velocities. For carbon-rich AGB stars this works well \citep{Winters00, Gautschy-Loidl04, nowotny10, nowotny11, Sacuto11}, but driving a wind by absorption of photons in an oxygen-rich environment is difficult (e.g. \citealt{Jeong03}). This is because the refractory dust grains in an oxygen-rich environment (e.g. olivine and alumina) are transparent in the optical, meaning they absorb few photons, and thus momentum, from the radiation field of the central star (which has a temperature of $\sim$3000 K). A suggestion to drive winds around oxygen-rich AGBs that seems to work well is momentum transfer to the grains by scattering \citep{hofner08, bladh12}. For this to be efficient, the grains need to grow to a size of $\sim$0.5 \mic. Since forsterite is a refractory solid, it could be a driver of the outflow of massive AGB stars. We tested if the grain size dependence of the scattering efficiency of forsterite grains might be a way of grain-size-filtering, which might explain the lower limit for the grain size of forsterite around \source.

The radiative transport code MCMax that we use for the models (see sect.~\ref{sec:model}) lets us calculate the $\beta$ of grains. This $\beta$ is the ratio of radiation pressure to gravitational force on the grain. The code takes into account the anisotropic scattering properties of the grains. A discussion on the driving of a superwind or a hyperwind is outside the scope of this paper, but we simply want to see how the $\beta$ of forsterite behaves as a function of grain size. Therefor we calculated the $\beta$ for forsterite grains at different grain sizes, shown in Fig. \ref{figBetas}. We see, as shown before by \cite{hofner08}, that the momentum transfer by scattering becomes more efficient when the grains approach a size of $\sim$0.5 \mic, but a maximum $\beta$ is also reached at 0.5 \mic and the scattering becomes less efficient at larger grain sizes. This is due to the decrease in the area to volume ratio of the grains. This shows that the lower-limit we find for the forsterite grains around \source is probably not directly linked to the wind driving by scattering on micron-sized grains. One could imagine that forsterite is the wind driving dust species and that a population of $\sim$0.5 \mic grains are blown out and that those subsequently grow to the observed size of 2-6 \mic, but since the density quickly drops after the onset of the outflow, this would require very special circumstances.

Another possible scenario to explain the lower-limit of the forsterite grain size is that an initial population of forsterite grains is formed with a size distribution ranging from sub-micron grains up to $\sim$6 \mic, but that the smaller grains are preferentially destroyed or amorphisised. If the smaller forsterite grains are preferentially destroyed, one has to explain why the small amorphous grains are not destroyed along with them. 
Instead of a destruction mechanism for both the amorphous and forsterite dust one could consider the selective amorphisation of forsterite grains instead. 

One possible amorphisation mechanism is the destruction of the crystalline lattice structure by ion bombardments (for example Ar$^{2+}$ or Fe$^{2+}$, \cite{borg80, Day77, Bradley94,Demyk01, Carrez02, jager03b, Brucato04, kemper04, kemperErr05}). \cite{kemper04} shows that for the interstellar medium (ISM) it takes several million years to destroy several percentage points by mass of sub-micron forsterite grains. Since the super- and hyperwind of \source combined would not even take more than several thousand years, these amorphisation mechanisms are too slow. Possibly, more efficient amorphisation can be obtained at the inner-rim of the AGB (hyper) wind, where the fast stellar wind will impact the recently stopped hyperwind \citep{Balick02}, but that could not easily explain the absence of small forsterite grains throughout the whole super and hyper outflow.


\section{Conclusions}
\label{sec:conc}
From studying the spectral features of forsterite in the spectrum of \source we can draw the following conclusions:
\begin{itemize}
\item The forsterite crystals in the outflow are dominated by a grain size population between 2 \mic and 6 \mic.
\item For the forsterite component we cannot distinguish between the \textit{Whole}, \textit{Superwind} or \textRefTwee{\textit{AGB-wind}} models. 
\item We exclude the \textit{Torus} and \textit{Disk} models.
\item The lower-limit for the forsterite grain size is likely unrelated to any wind driving mechanism based on scattering on micron-sized dust grains
\item The blue shoulder of the 69 \mic band cannot be explained by cold forsterite. We suggest the possibility of micron-sized ortho-enstatite as a carrier.
\end{itemize}

We also speculate that the following processes might be related to the hyperwind:
\begin{itemize}
\item The amorphous grains grow to several times 100 \mic in size during the hyperwind
\item Forsterite grains can grow up to sizes as large as $\sim$6 \mic during the hyperwind
\item The absence of sub-micron forsterite grains could be due to an amorphisation mechanism currently unknown
\end{itemize}

\bibliographystyle{aa}
\bibliography{references}

\end{document}